\pdfoutput=1
\documentclass[iop,onecolumn,showpacs,preprintnumbers,superscriptaddress,amsmath,amssymb]{revtex4-2}
\usepackage{graphicx}
\usepackage{float}
\usepackage{dcolumn}
\usepackage{color}
\usepackage[dvipsnames]{xcolor} 
\usepackage{latexsym,bm}
\usepackage[normalem]{ulem}
\usepackage{multirow}
\usepackage{appendix}
\usepackage{amsmath}
\usepackage{amsfonts}
\usepackage{booktabs}
\usepackage{subcaption}
\usepackage{tabularx}

\usepackage{threeparttable} 

\newcommand{\MC}[1]{\textcolor{black}{{#1}}}

\newcommand{\SL}[1]{\textcolor{black}{{#1}}}
\newcommand{\SSL}[1]{\textcolor{black}{{#1}}}

\newcommand{\tx}[1]{\text{#1}}

\newcommand{\kernel}[1]{w^{( #1 )}(\mathbf{r}-\mathbf{r^{\prime}})}
\newcommand{\cpn}[1]{CPN$_#1$}
\begin{document}

\hyphenpenalty=5000
\tolerance=1000

\title{Multi-channel machine learning based nonlocal kinetic energy density functional for semiconductors}

\author{Liang Sun}
\affiliation{HEDPS, CAPT, School of Physics and College of Engineering, Peking University, Beijing 100871, P. R. China}
\author{Mohan Chen}
\email{mohanchen@pku.edu.cn}
\affiliation{HEDPS, CAPT, School of Physics and College of Engineering, Peking University, Beijing 100871, P. R. China}
\affiliation{AI for Science Institute, Beijing 100080, P. R. China}
\date{\today}

\begin{abstract}
\SL{
The recently proposed machine learning-based physically-constrained nonlocal (MPN) kinetic energy density functional (KEDF) can be used for simple metals and their alloys [Phys. Rev. B \textbf{109}, 115135 (2024)].
However, the MPN KEDF does not perform well for semiconductors.
Here we propose a multi-channel MPN (CPN) KEDF, which extends the MPN KEDF to semiconductors by integrating information collected from multiple channels, with each channel featuring a specific length scale in real space. 
The CPN KEDF is systematically tested on silicon and binary semiconductors.
We find that the multi-channel design for KEDF is beneficial for machine-learning-based models in capturing the characteristics of semiconductors, particularly in handling covalent bonds.
In particular, the \cpn{5} KEDF, which utilizes five channels, demonstrates excellent accuracy across all tested systems.
These results offer a new path for generating KEDFs for semiconductors.
}
\end{abstract}
\maketitle

\section{Introduction}
\SL{
The Kohn-Sham formulation of density functional theory (KSDFT)~\cite{64PR-Hohenberg, 65PR-Kohn} is well balanced in terms of accuracy and computational efficiency for systems comprising up to several hundred atoms.
However, in general, KSDFT needs Kohn-Sham orbitals to calculate the non-interacting kinetic energy $T_s$ of electrons. In this regard, a diagonalization step for the Hamiltonian is needed to yield the Kohn-Sham orbitals and corresponding eigenvalues. Thus, the application of KSDFT to larger systems is limited by its $O(N^3)$ computational complexity, where $N$ is the number of atoms. On the other hand, orbital-free density functional theory (OFDFT)~\cite{02Carter,18JMR-Witt, 23CR-Mi} circumvents the need for Kohn-Sham orbitals by calculating $T_s$ directly from the charge density $\rho(\mathbf{r})$, and OFDFT does not need the sampling of Brillouin zone, thereby achieving a more affordable computational complexity, typically $O(N\ln{N})$ or $O(N)$.~\cite{08CPC-Ho-PROFESS, 10CPC-Hung-PROFESS, 15CPC-Chen-PROFESS, 16CPC-Mi-atlas}
Importantly, the approximated form of the kinetic energy density functional (KEDF) in OFDFT largely determines its accuracy. Unfortunately, the exact form for the KEDFs for systems like semiconductors is largely unknown. Given that $T_s$ is comparable in magnitude to the total energy, developing accurate KEDFs has been challenging in the past few decades.
}

\SL{
Despite formidable challenges, there has been a continuous focus on the advancement and refinement of analytical KEDFs in recent years.~\cite{12CPC-Karasiev, 18JMR-Witt, 24WIRCMS-Xu}
The local/semilocal and nonlocal functionals are the two main forms of KEDFs.
Generally speaking, semilocal KEDFs are recognized for their computational efficiency, whereas nonlocal KEDFs tend to provide superior accuracy.
Specifically, the kinetic energy density of local/semilocal KEDF is designed as a function of quantities such as the charge density, the charge density gradient, the Laplacian of charge density, or higher-order derivatives of the charge density.~\cite{27-Thomas-local, 27TANL-Fermi-local, 35-vW-semilocal, 18B-Luo-semilocal, 18JPCL-Constantin-semilocal, 20Kang-semilocal, 24JCTC-Wang-semilocal}
Meanwhile, the kinetic energy density within nonlocal KEDFs is formulated as a functional of charge density.~\cite{92B-Wang-nonlocal, 99B-Wang-nonlocal, 10B-Huang-nonlocal, 18JCP-Mi-nonlocal, 21B-Shao-nonlocal, 23B-Sun-TKK, 24JCP-Bhattacharjee-nonlocal} Interestingly, nonlocal KEDFs based on the Lindhard response function are mainly suitable for simple metals,~\cite{92B-Wang-nonlocal, 99B-Wang-nonlocal} while the KEDFs proposed to describe semiconductor systems do not perform well for simple metals.~\cite{10B-Huang-nonlocal, 18JCP-Mi-nonlocal, 21B-Shao-nonlocal} Consequently, an analytical KEDF that is effective for both simple metal and semiconductor systems is still lacking.
}

\SL{
Another approach to constructing KEDFs is through the use of machine learning (ML) techniques, which have been widely involved in the developments of computational physics,~\cite{18prl-zhang-dp, 20CPC-Zhang-dpgen, 20JCTC-chen-deepks, 21L-Kasim-mlxc, 21s-kirkpatrick-dm21, 22PRR-nagai-mlxc, 22NCS-Li-deepH, 23Science-Huang-mlfp, 23B-Lv-deepcharge, 24JCTC-Shakiba, 24PRL-YangLi, 24B-Bystrom-mlxc} and several ML-based KEDFs have been proposed in recent years.~\cite{12L-Snyder-mlof, 18TJCP-Seino-mlof, 18TJCP-Hollingsworth-mlof, 19CPL-Seino-mlof, 20JCTC-Meyer-mlof, 21PRR-Imoto-mlof, 22JCTC-Ryczko-mlof, 23JCTC-Pablo-mlof, 24NCS-Zhang-mlof, 24B-Sun-mlof}
In particular, several local/semilocal ML-based KEDFs have surpassed the accuracy of traditional semilocal KEDFs; however, due to the lack of nonlocal characteristics of $T_s$, they can scarcely match the accuracy of nonlocal KEDFs.~\cite{18TJCP-Seino-mlof, 21PRR-Imoto-mlof}
To introduce nonlocal information, some KEDFs project the charge density onto a set of bases and use the projection coefficients as descriptors~\cite{24NCS-Zhang-mlof}, while others employ convolution methods, either to define the descriptors~\cite{24B-Sun-mlof} or to directly utilize convolutional neural networks~\cite{22JCTC-Ryczko-mlof}.
Furthermore, some KEDFs incorporate the positions of atoms to enhance accuracy.~\cite{19CPL-Seino-mlof, 24NCS-Zhang-mlof}
Nevertheless, further development is still needed for an accurate, transferable, and computationally stable ML-based kinetic energy density functional.}

\SL{
Recently, based on the deep neural network, the ML-based physical-constrained nonlocal KEDF (MPN KEDF)~\cite{24B-Sun-mlof} was proposed for simple metals and their alloys.
Four descriptors are designed as the inputs for the neural network in the MPN KEDF, with one being a semilocal descriptor and the other three capturing the nonlocal information.
In addition, by carefully designing the output of the neural network and the post-processing, the MPN KEDF adheres to three exact physical constraints: the scaling law, the free electron gas (FEG) limit, and the non-negativity of the Pauli energy density.
The MPN KEDF was validated for simple metals, including Li, Mg, Al, and 59 alloys.
Consequently, the new KEDF exhibits high accuracy and transferability in all these tests, outperforming semilocal KEDFs and approaching the accuracy of nonlocal KEDFs.
This work demonstrates that incorporating nonlocal information and adhering to exact physical constraints are critical for improving the accuracy, transferability, and stability of ML-based KEDFs. 
Nevertheless, the accuracy of the MPN KEDF for semiconductor systems is still unsatisfactory ({\it vide infra}).
}

\SL{
Based on the previous work, here we construct a multi-channel ML-based physical-constrained nonlocal KEDF (CPN KEDF) that extends the MPN KEDF to semiconductor systems.}
The proposed CPN KEDF integrates information collected from several channels, with each channel capturing features among specific scale.
The CPN KEDF with $n$ channels is labeled as \cpn{n} KEDF.
As an intermediate step towards constructing an ML-based KEDF that is effective for both simple metals and semiconductor systems, we focus on semiconductor systems in this work.
The performance of the CPN KEDF is systematically evaluated on a series of semiconductor systems, including silicon (Si) and binary semiconductors, \SL{which serve as standard benchmarks in the development of KEDFs,}~\cite{10B-Huang-nonlocal, 12B-Xia-nonlocal, 14JCP-Shin-nonlocal, 18B-Constantin-nonlocal, 21B-Shao-nonlocal} and the \cpn{5} KEDF exhibits excellent accuracy in above systems.
\SL{Furthermore, other semiconductors such as germanium were not included in this study because the corresponding local pseudopotential is not available.}
Notably, mixing features of different scales is crucial for ML-based models to capture the characteristics of semiconductors and to handle covalent bonds.

%
%
%
%

\begin{figure}[t]
	\centering
	\includegraphics[width=\linewidth]{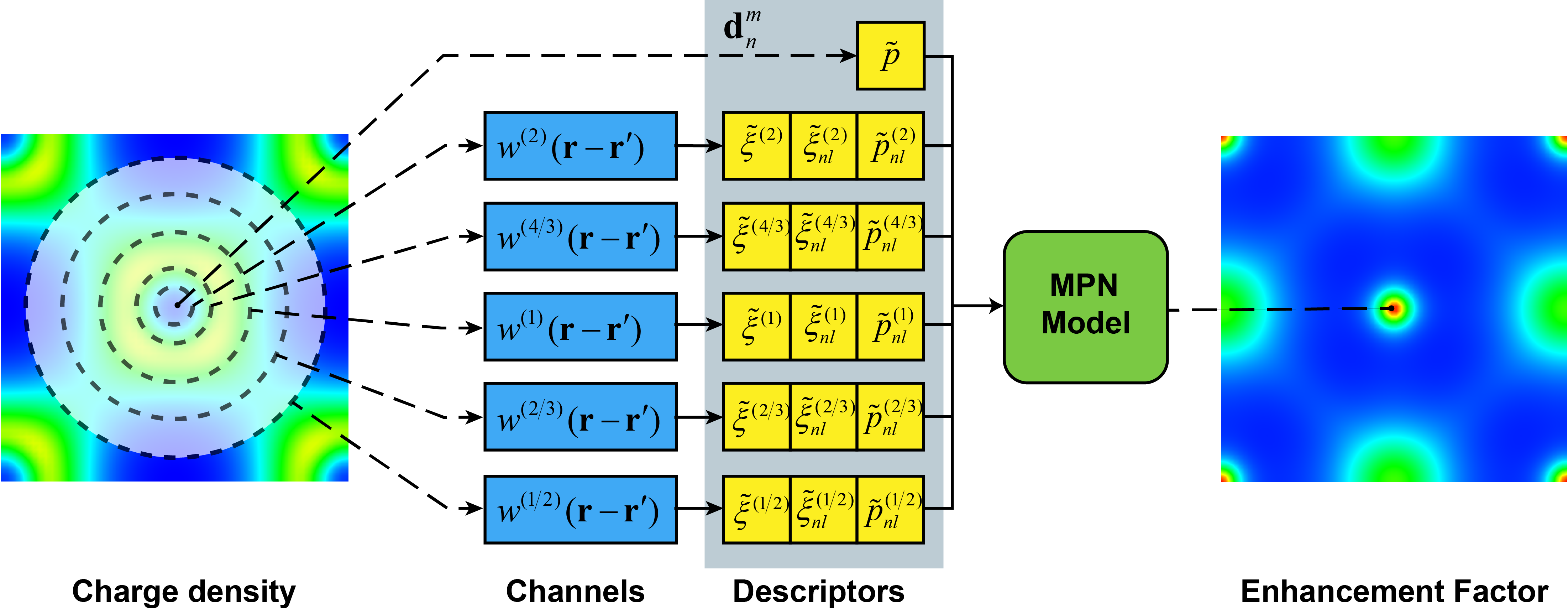}\\
	\caption{
Workflow of the CPN KEDF. 
Initially, the charge density is introduced into several channels, each characterized by a specific kernel function $\kernel{\lambda} \equiv \lambda^3 W(\lambda(\mathbf{r}-\mathbf{r^{\prime}}))$.
Each channel generates three nonlocal descriptors $\Tilde{\xi}^{(\lambda)}$, $\Tilde{\xi}^{(\lambda)}_{nl}$, and $\Tilde{p}^{(\lambda)}_{nl}$.
Subsequently, these descriptors are gathered and packaged into a vector $\mathbf{d}^{m}_n$, and $\mathbf{d}^{m}_n$ is input into the MPN model as proposed by Ref.~\onlinecite{24B-Sun-mlof}, which strictly adheres to three physical constraints: the scaling law of $T_\theta$, the FEG limit, as well as the non-negativity of $\tau_\theta$.
Ultimately, the enhancement factor of the Pauli energy $F_{\rm{\theta}}^{\rm{NN}}$ is obtained.
	}\label{fig:workflow}
\end{figure}


The paper is organized as follows. \SL{In Section II, we present an ML-based KEDF with a multi-channel architecture and detail the numerical aspects of the KSDFT and OFDFT calculations.
The performances of the CPN KEDF and results are discussed in Section III.}
Finally, the conclusions are drawn in Section IV.

\section{Method}
\subsection{Pauli Energy and Pauli Potential}

In the field of OFDFT, the Pauli energy~\cite{88PRA-Levy-pauli}, introduced by the Pauli exclusion principle, is defined as
\begin{equation}
    T_{\rm{\theta}} = T_{{s}} - T_{\rm{vW}},
\end{equation}
where $T_s$ is non-interacting kinetic energy, and  
\begin{equation}
T_{\rm{vW}} = \frac{1}{8} \int {\frac{{\left| {\nabla \rho ({\mathbf{r}})} \right|}^2}{\rho \,({\mathbf{r}})} \,\tx{d}^3 {\mathbf{r}}}
\end{equation}
is the von Weizs$\mathrm{\Ddot{a}}$cker (vW) KEDF,~\cite{35-vW-semilocal} the rigorous lower bound to the $T_s$, with $\rho(\mathbf{r})$ being the charge density. 

In general, \MC{the} Pauli energy takes the form of
\begin{equation}
    T_{\rm{\theta}} = \int{\tau_{\theta}({\mathbf{r}}) {\rm{d}}^3 {\mathbf{r}}} = \int{\tau_{\rm{TF}}({\mathbf{r}}) F_{\rm{\theta}}({\mathbf{r}}) {\rm{d}}^3 {\mathbf{r}}},
\end{equation}
where $\tau_\theta({\mathbf{r}})$ denotes the Pauli energy density, the Thomas-Fermi (TF) kinetic energy density~\cite{27-Thomas-local, 27TANL-Fermi-local} \SSL{$\tau_{\rm{TF}}({\mathbf{r}})$} is
\begin{equation}
\tau_{\rm{TF}}({\mathbf{r}}) = \frac{3}{10}(3\pi^2)^{2/3} \rho^{5/3}({\mathbf{r}}),
\end{equation}
and $F_{\rm{\theta}}({\mathbf{r}})$ \SSL{represents} the enhancement factor.
The corresponding Pauli potential is \SSL{defined as} 
\begin{equation}
V_{\rm{\theta}}(\mathbf{r}) = \delta T_{\rm{\theta}}/\delta \rho(\mathbf{r}).
\end{equation}

\SL{The Pauli energy and the associated Pauli potential are subject to several exact physical constraints.} First, the scaling law $T_{\rm{\theta}}[\rho_{\lambda}] = \lambda^2 T_{\rm{\theta}}[\rho]$, where $\rho_{\lambda}=\lambda^3\rho(\lambda \mathbf{r})$ with $\lambda$ being a positive number.~\cite{88PRA-Levy-pauli}.
Second, in the FEG limit, the Pauli energy and Pauli potential revert to the TF model, which requires $F_{\rm{\theta}}(\mathbf{r})|_{\rm{FEG}} = 1$ and $\delta F_\theta(\mathbf{r})/\delta \rho(\mathbf{r^{\prime}})|_{\rm{FEG}} = 0$.
Third, the non-negativity constraint ensures $F_{\rm{\theta}}(\mathbf{r}) \geq 0$.

In order to gather the training data, the Pauli energy and Pauli potential data for a set of selected systems were calculated using KSDFT.
Specifically, in a spin degenerate system, the Pauli energy density \SSL{and Pauli potential} can be analytically expressed as~\cite{88PRA-Levy-pauli}
\begin{equation}
    \tau_{\theta}^{\rm{KS}}(\mathbf{r}) = \sum_{i=1}^M {f_i|\nabla \psi_i(\mathbf{r})|^2} - \frac{|\nabla\rho(\mathbf{r})|^2}{8\rho(\mathbf{r})},
    \label{eq.pauli_e}
\end{equation}
\begin{equation}
    V_{\theta}^{\rm{KS}}(\mathbf{r}) = \rho^{-1}(\mathbf{r}) \left( \tau_{\theta}^{\rm{KS}}(\mathbf{r}) +2 \sum_{i=1}^M {f_i(\varepsilon_M-\varepsilon_i)\psi_i^*(\mathbf{r})\psi_i(\mathbf{r})}\right),
    \label{eq.pauli_p}
\end{equation}
where $\psi_i(\mathbf{r})$ \SSL{represents} an occupied Kohn-Sham orbital with index $i$, $\varepsilon_i$ and $f_i$ are the corresponding eigenvalue and occupied number, respectively.
\SSL{Furthermore}, $M$ \SSL{denotes} the highest occupied state, and $\varepsilon_M$ is the eigenvalue of $\psi_M(\mathbf{r})$, \SL{which corresponds to the chemical potential.}

\subsection{Multi-channel MPN model}
\SL{The charge densities of simple metals and semiconductors exhibit significant differences.
While the charge density of simple metals is close to that of a free electron gas, the charge density of semiconductors consists of two distinct parts: highly localized electrons in the covalent bond regions and delocalized electrons.
One approach to handle this complexity is the density-decomposed method~\cite{12B-Xia-nonlocal}, which calculates the contributions from these two parts using different KEDFs.
However, physically distinguishing these two parts is a challenging task, and this method also suffers from computational instability.
In this work, we propose a multi-channel architecture to address this challenge.
We compressed the kernel function utilized by the MPN KEDF to capture the features of highly localized electrons and stretched the kernel function to capture the features of delocalized density. 
By integrating these features across different scales, we extend the MPN KEDF to the CPN KEDF.}

\SL{Figure~\ref{fig:workflow} illustrates the workflow of the CPN KEDF.}
Firstly, in order to capture the information among different scales, the kernel function $W(\mathbf{r}-\mathbf{r^{\prime}})$ utilized by the MPN KEDF~\cite{24B-Sun-mlof} is scaled to $\kernel{\lambda} \equiv \lambda^3 W(\lambda(\mathbf{r}-\mathbf{r^{\prime}}))$ with $\lambda$ being a positive number.
As depicted in figure~\ref{fig:workflow} and figure~\ref{fig:si_curve}(b), the kernel functions with larger values of $\lambda$ are more localized and thus cover smaller areas.
Each $\kernel{\lambda}$ with a typical $\lambda$ defines a ``channel'', which transforms the input charge density into three nonlocal descriptors $\{\Tilde{p}^{(\lambda)}_{\rm{nl}}, \Tilde{\xi}^{(\lambda)}, \Tilde{\xi}^{(\lambda)}_{\rm{nl}}\}$.
In detail, the nonlocal descriptors are defined as
\begin{equation}
    \begin{aligned}
            \Tilde{\xi}^{(\lambda)}(\mathbf{r}) &= \tanh{\Big(\chi_{\xi^{(\lambda)}}\xi^{(\lambda)}(\mathbf{r})\Big)},\\
            \Tilde{\xi}^{(\lambda)}_{\rm{nl}}(\mathbf{r}) &= \int{\kernel{\lambda}\Tilde{\xi}^{(\lambda)}(\mathbf{r^{\prime}}){\rm{d}}^3 \mathbf{r^{\prime}}},\\
            \Tilde{p}^{(\lambda)}_{\rm{nl}}(\mathbf{r}) &= \int{\kernel{\lambda}\Tilde{p}(\mathbf{r^{\prime}}){\rm{d}}^3 \mathbf{r^{\prime}}},
    \end{aligned}
\end{equation}
with $\xi^{(\lambda)}(\mathbf{r})=\frac{\int{\kernel{\lambda}\rho^{1/3}(\mathbf{r^{\prime}}){\rm{d}}^3 \mathbf{r^{\prime}}}}{\rho^{1/3}(\mathbf{r})}$, and $\chi_{\xi^{(\lambda)}}$ is a hyper-parameter to control the data distribution.
Additionally, the semilocal descriptor $\Tilde{p}$ takes the form of
\begin{equation}
        \Tilde{p}(\mathbf{r}) = \tanh{\Big(\chi_p p(\mathbf{r})\Big)},
\end{equation}
with $p(\mathbf{r}) = |\nabla \rho(\mathbf{r})|^2 / \Big(2(3\pi^2)^{1/3} \rho^{4/3}(\mathbf{r})\Big)^2$ and $\chi_p = 0.2$.

Second, once the semilocal descriptor $\Tilde{p}$ and nonlocal descriptors from all channels on grid $\mathbf{r}$ are obtained, they are gathered and packaged into a vector $\mathbf{d}^{m}_n(\mathbf{r}) = \left(\Tilde{p}, \Tilde{p}^{(\lambda_1)}_{\rm{nl}}, \Tilde{\xi}^{(\lambda_1)}, \Tilde{\xi}^{(\lambda_1)}_{\rm{nl}}, \cdots, \Tilde{p}^{(\lambda_n)}_{\rm{nl}}, \Tilde{\xi}^{(\lambda_n)}, \Tilde{\xi}^{(\lambda_n)}_{\rm{nl}} \right)$, where $n$ denotes the number of channel, and $m$ denotes the number of descriptors.
Since each channel generates three nonlocal descriptors, including the semilocal descriptor $\Tilde{p}$, we have $m = 3n + 1$.
The descriptor vector $\mathbf{d}^m_n(\mathbf{r})$ is then fed into the MPN model, which predicts the enhancement factor $F_{\rm{\theta}}^{\rm{NN}}$ on the grid $\mathbf{r}$, ensuring the aforementioned three physical constraints: the scaling law, the FEG limit, and non-negativity, as discussed in Ref.~\onlinecite{24B-Sun-mlof}.

Finally, after obtaining the enhancement factor $F_{\rm{\theta}}^{\rm{NN}}$, the non-interacting electron kinetic energy is calculated through
\begin{equation}
    T_{\rm{CPN}_n} = T_{\rm{vW}} + \int{\tau_{\rm{TF}}(\mathbf{r})} F_{\rm{\theta}}^{\rm{NN}} \left(\mathbf{d}^m_n(\mathbf{r})\right) {\rm{d}}^3{\mathbf{r}},
\end{equation}
and corresponding potential is defined as $V_{\rm{CPN}_n}(\mathbf{r}) = \delta T_{\rm{CPN}_n}/\delta \rho(\mathbf{r})$.

In this work, in order to evaluate the validity of the multi-channel architecture, we have established three distinct models, each utilizing one, three, and five channels, respectively, and the detailed parameters are listed in table~\ref{tab:CPN}.
%
We note that the descriptors of \cpn{1} KEDF are identical to those of the MPN KEDF, so the performance of \cpn{1} represents the performance of the MPN KEDF.
In addition, the five kernel functions utilized by \cpn{5} KEDF are displayed in figure~\ref{fig:si_curve}(b).
These kernel functions tend to zero at different radii, enabling them to capture information across a range of specific scales, which are also illustrated in figure~\ref{fig:workflow}.

\begin{table*}[t]
	\centering
	\caption{
 Detailed parameters of \cpn{1}, \cpn{3}, and \cpn{5} KEDFs, including the descriptor vectors $\mathbf{d}^m_n$, the scaling factor $\lambda$, as well as the corresponding hyper-parameters $\chi_{\xi^{(\lambda)}}$ for the nonlocal descriptors $\Tilde{\xi}^{(\lambda)}$ and $\Tilde{\xi}^{(\lambda)}_{\rm{nl}}$.
        }
	\begin{tabularx}{0.9\linewidth}{
			>{\raggedright\arraybackslash\hsize=0.5\hsize\linewidth=\hsize}X
			>{\raggedright\arraybackslash\hsize=2\hsize\linewidth=\hsize}X
			>{\raggedright\arraybackslash\hsize=0.5\hsize\linewidth=\hsize}X}
		\hline\hline
KEDF		&Descriptors 	&$\chi_{\xi^{(\lambda)}}$\\
            \hline
\cpn{1} &
$
\mathbf{d}^4_1 = \left(\Tilde{p}, \Tilde{p}^{(1)}_{\rm{nl}}, \Tilde{\xi}^{(1)}, \Tilde{\xi}^{(1)}_{\rm{nl}}\right)
$
&$\{1\}$\\
\cpn{3} &
$
\mathbf{d}^{10}_3 = \left(\Tilde{p}, \Tilde{p}^{(2)}_{\rm{nl}}, \Tilde{\xi}^{(2)}, \Tilde{\xi}^{(2)}_{\rm{nl}}, \Tilde{p}^{(1)}_{\rm{nl}}, \Tilde{\xi}^{(1)}, \Tilde{\xi}^{(1)}_{\rm{nl}}, \Tilde{p}^{(1/2)}_{\rm{nl}}, \Tilde{\xi}^{(1/2)}, \Tilde{\xi}^{(1/2)}_{\rm{nl}}\right)
$  &$\{3, 1, 0.6\}$\\
\cpn{5} &$
\begin{aligned}
\mathbf{d}^{16}_5 = \big(&\Tilde{p}, \Tilde{p}^{(2)}_{\rm{nl}}, \Tilde{\xi}^{(2)}, \Tilde{\xi}^{(2)}_{\rm{nl}}, \Tilde{p}^{(4/3)}_{\rm{nl}}, \Tilde{\xi}^{(4/3)}, \Tilde{\xi}^{(4/3)}_{\rm{nl}}, \Tilde{p}^{(1)}_{\rm{nl}}, \Tilde{\xi}^{(1)}, \Tilde{\xi}^{(1)}_{\rm{nl}}, 
\\
&\Tilde{p}^{(2/3)}_{\rm{nl}}, \Tilde{\xi}^{(2/3)}, \Tilde{\xi}^{(2/3)}_{\rm{nl}},  \Tilde{p}^{(1/2)}_{\rm{nl}}, \Tilde{\xi}^{(1/2)}, \Tilde{\xi}^{(1/2)}_{\rm{nl}}\big)
\end{aligned}
$  &$\{3, 1.5, 1,0.6,  0.6\}$\\
		\hline\hline
	\end{tabularx}
	\label{tab:CPN}
\end{table*}

\SL{The neural network (NN) of CPN KEDF consists of an input layer with $m$ neurons, three hidden layers each containing 100 neurons, and an output layer with a single neuron.
The activation functions employed in the hidden layers are selected to be hyperbolic tangent functions.}

We take the same definition of the loss function as the MPN KEDF, which has been proven to work well.~\cite{24B-Sun-mlof} 
It takes the form of
\begin{equation}
        L=\frac{1}{N}\sum_{\mathbf{r}}{\left[ \left(\frac{F_\theta^{\rm{NN}}- F^{\rm{KS}}_{\theta}}{\bar{F}^{\rm{KS}}_{\theta}}\right)^2 +
      \left(\frac{V_\theta^{\rm{CPN}} - V^{\rm{KS}}_{\theta}}{\bar{V}^{\rm{KS}}_{\theta}}\right)^2 \right]}
      + \left[F^{\rm{NN}}|_{\rm{FEG}}-\ln(e-1)\right]^2.
      \label{eq.loss}
\end{equation}
Where $N$ is the number of grid points, and $\bar{F}^{\rm{KS}}_{\theta}$ ($\bar{V}^{\rm{KS}}_{\theta}$) represents the mean of $F^{\rm{KS}}_{\theta}$ ($V^{\rm{KS}}_{\theta}$).

\SL{The training set comprises ten semiconductor structures, including cubic diamond (CD) Si and nine cubic zincblende (ZB) semiconductors: AlP, AlAs, AlSb, GaP, GaAs, GaSb, InP, InAs, and InSb.
To collect the ground state charge density, Pauli energy density, Pauli potential, and the relevant descriptors, KSDFT calculations were performed.
These calculations utilized a $27\times27\times27$ grid,  resulting in 196,830 grid points for the entire training set.}

\subsection{Numerical Details}

\SL{We conducted both OFDFT and KSDFT calculations using the ABACUS 3.6.5 software package.~\cite{16Li-CMS-ABACUS}
For OFDFT calculations employing the Wang-Govind-Carter (WGC) KEDF~\cite{99B-Wang-nonlocal} and Huang-Carter (HC) KEDF~\cite{10B-Huang-nonlocal}, the PROFESS 3.0 package was used.~\cite{15CPC-Chen-PROFESS}
The CPN KEDF was implemented within ABACUS with the assistance of the libtorch library,~\cite{19ANIPS-paszke-pytorch} while the libnpy library was employed to dump the data.}
The plane-wave energy cutoffs for KSDFT calculations were set to 800 eV, while those for OFDFT were equivalently set to 3200 eV.
The Monkhorst-Pack $k$-point samplings~\cite{76B-Monkhorst} utilized in KSDFT ensure that the spacing between $k$ points does not exceed 0.05 Bohr$^{-1}$.
For both OFDFT and KSDFT calculations, \SL{ the Perdew-Burke-Ernzerhof (PBE) exchange-correlation functional~\cite{96PRL-Perdew-PBE} and bulk-derived local pseudopotentials (BLPS)~\cite{08PCCP-Huang-BLPS} were adopted}.

\SL{Several analytical KEDFs were utilized for comparative analysis, including semilocal KEDFs, such as the TF$\rm{\lambda}$vW~\cite{83pra-berk-semilocal} and the Luo-Karasiev-Trickey (LKT) KEDFs~\cite{18B-Luo-semilocal}, as well as nonlocal KEDFs such as the WGC and HC KEDFs.
For the TF$\rm{\lambda}$vW KEDF, the parameter $\rm{\lambda}$ was set to 0.2.
In the case of the LKT KEDF, the parameter $a$ was configured to 1.3, following the original study~\cite{18B-Luo-semilocal}.
Additionally, for the WGC KEDF, the parameters were configured as follows: $\alpha$=$\frac{5+\sqrt{5}}{6}$, $\beta$=$\frac{5-\sqrt{5}}{6}$ and $\gamma$=4.2}~\cite{08CPC-Ho-PROFESS}.
As for the HC KEDF, the optimal values of $\gamma$ and $\beta$ were taken for each structure.~\cite{10B-Huang-nonlocal}

\SL{The Mean Absolute Relative Error (MARE) for a property $x$ is defined as follows: 
\begin{equation}
    {\rm{MARE}}=\frac{1}{N}\sum_i^N{\left|\frac{x_i^{\rm{OF}} - x_i^{\rm{KS}}}{x_i^{\rm{KS}}}\right|},\\
    \label{eq.mare}
\end{equation}
where $N$ is the total number of data points, and $x_i^{\rm{OF}}$ and $x_i^{\rm{KS}}$ represent the values of the property obtained from OFDFT and KSDFT calculations, respectively.}

\section{Results and discussion}

\SL{
A test set comprising ten semiconductors was established to evaluate the accuracy and transferability of the CPN KEDF.
This test set includes the equilibrium structures of hexagonal diamond (HD) Si and nine (hexagonal) wurtzite (WZ) semiconductors (AlP, AlAs, AlSb, GaP, GaAs, GaSb, InP, InAs, InSb) as determined by KSDFT.
We calculated the ground state energies and charge densities of the structures in both the training and test sets using various kinetic energy density functionals (KEDFs).
These results were then systematically compared with those obtained from KSDFT calculations.
}

\begin{figure*}[t]
    \centering

    \begin{subfigure}{0.49\textwidth}
    \centering
    \includegraphics[width=0.95\linewidth]{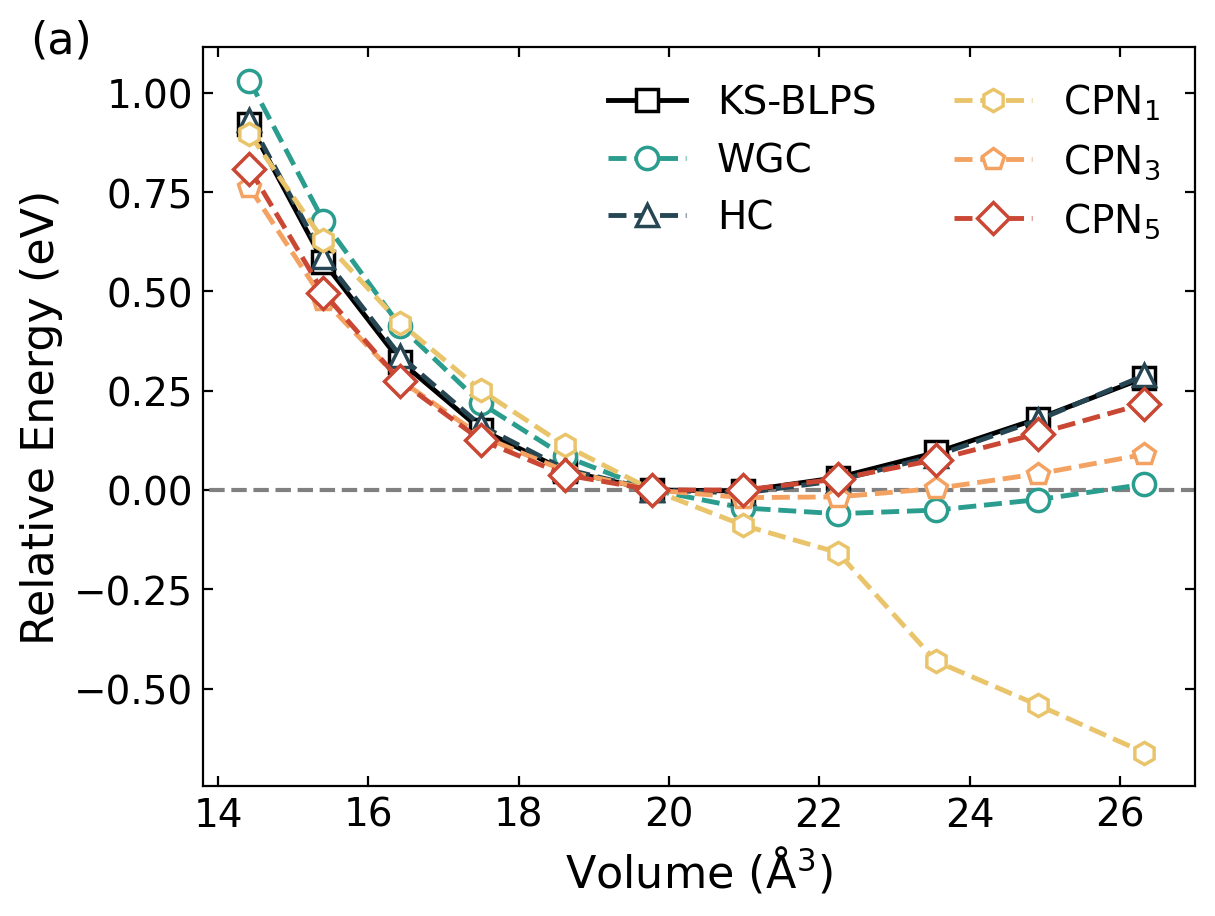}
    \label{fig:si_e_v}
    \end{subfigure}
    \begin{subfigure}{0.5\textwidth}
    \centering
    \includegraphics[width=0.97\linewidth]{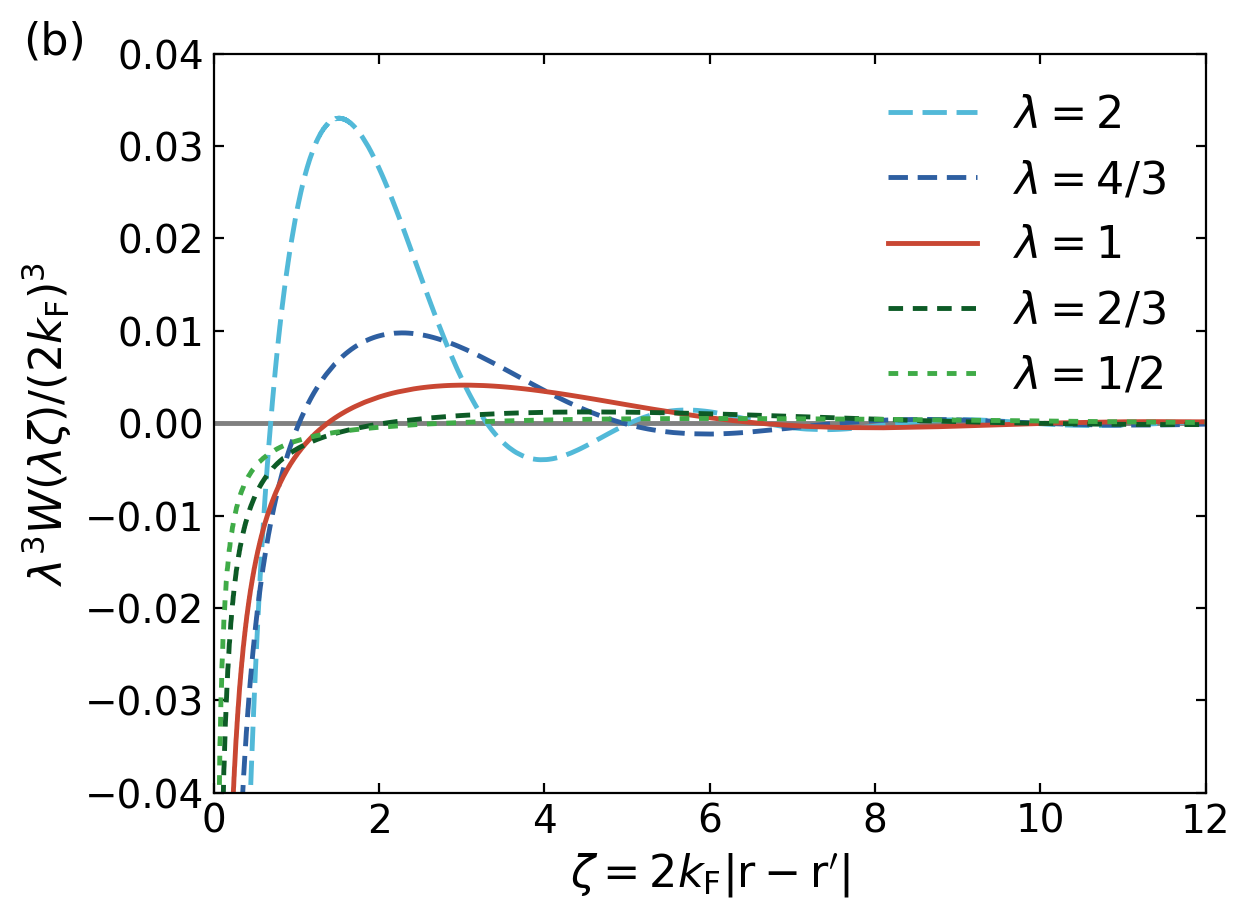}
    \label{fig:kernels}
    \end{subfigure}

    \caption{
    (a) Energy-Volume curve of cubic diamond (CD) Si as obtained by the WGC, HC, \cpn{1}, \cpn{3}, \cpn{5} KEDFs, and KSDFT with BLPS.
    Note that all the curves are shifted by subtracting the equilibrium energy obtained by KSDFT.
    (b) The kernel functions $w^{(\lambda)}(\zeta) \equiv \lambda^3 W(\lambda\zeta)$ in real space, with scaling factor $\lambda = 2, 4/3, 1, 2/3,$ and $1/2$. 
    Here $W(r-r')$ is the kernel function utilized by the MPN KEDF, and $k_{\rm{F}} = (3\pi^2\rho_0)^{1/3}$ is the Fermi vector with $\rho_0$ being the average charge density.
    Note that the kernel functions displayed here are divided by $(2k_{\rm{F}})^3$ to render them dimensionless.
    }
    \label{fig:si_curve}
\end{figure*}

First, we examine the performance of CPN KEDFs in the training set.
Figure~\ref{fig:si_curve}(a) displays the energy-volume curve of CD Si obtained by WGC, HC, \cpn{1}, \cpn{3}, and \cpn{5} KEDFs, as well as KSDFT.
The HC KEDF, designed specifically for semiconductors, produces a curve nearly identical to that of KSDFT, while the WGC KEDF yields a much poorer result.
Regarding the CPN KEDFs, the \cpn{1} KEDF fails to produce a smooth energy-volume curve.
However, as the number of channels increases, the accuracy of the CPN KEDF improves significantly.
For instance, the \cpn{3} KEDF exceeds the WGC KEDF by providing a curve closer to that of KSDFT, and the \cpn{5} KEDF approaches the accuracy of both the HC KEDF and KSDFT.
As a result, the multi-channel architecture enhances the CPN KEDF's ability to capture the features of CD Si, thereby improving both the accuracy and stability of the CPN KEDF.

\begin{figure*}[t]
    \centering
    
    \begin{subfigure}{0.49\textwidth}
    \centering
    \includegraphics[width=0.95\linewidth]{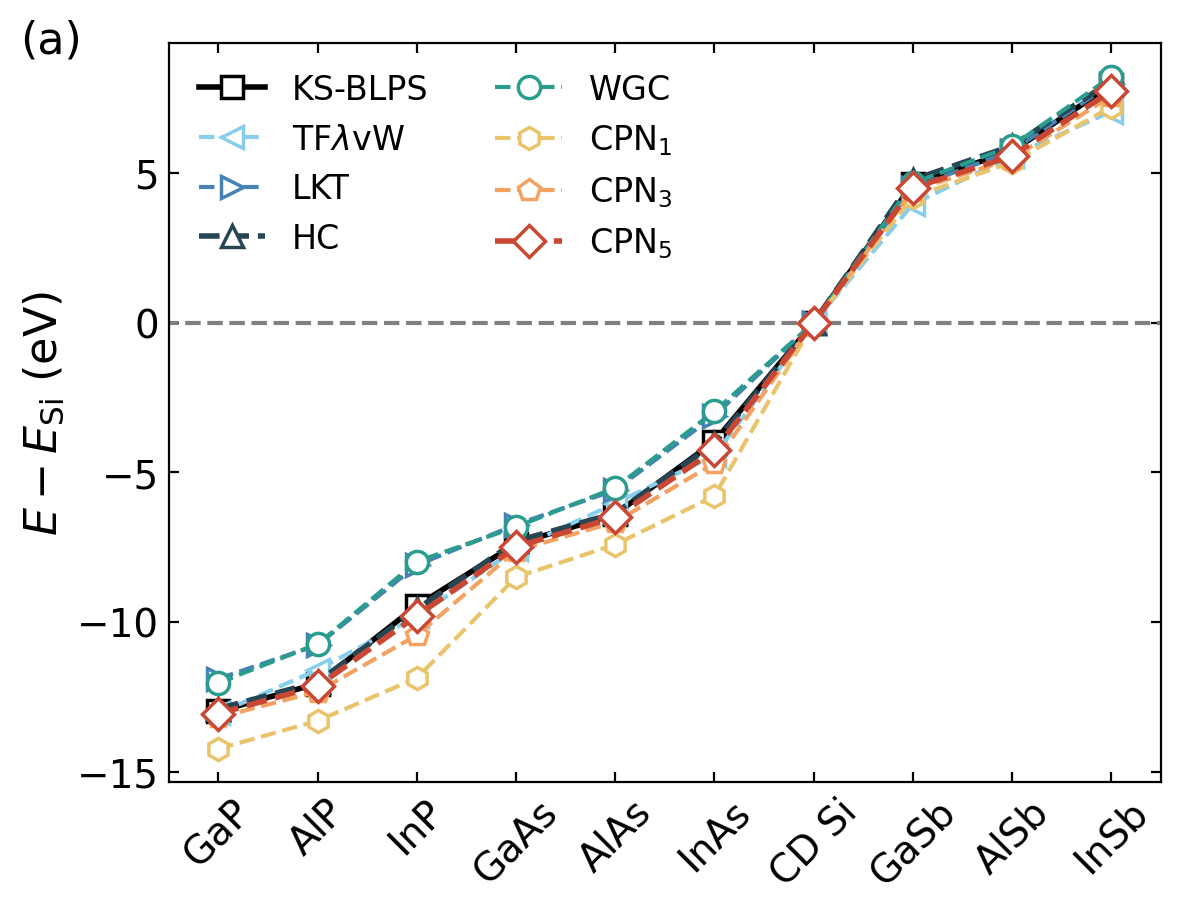}
    \label{fig:E_diff_cd}
    \end{subfigure}
    \begin{subfigure}{0.49\textwidth}
    \centering
    \includegraphics[width=0.95\linewidth]{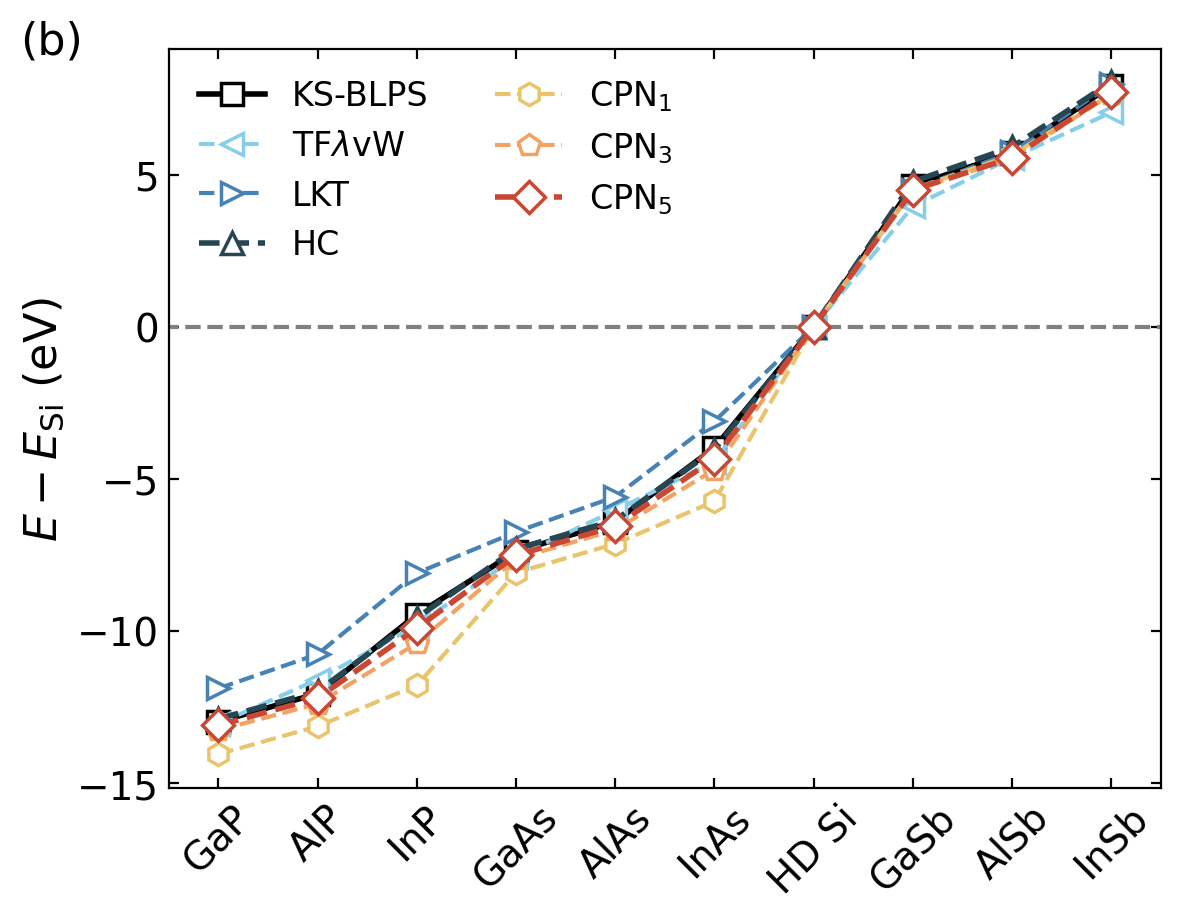}
    \label{fig:E_diff_hd}
    \end{subfigure}
    
    \caption{
    (a) Relative energy differences between CD Si and ZB semiconductors as obtained by OFDFT and KSDFT, where $E$ denotes the energy of specific structure, and $E_{\rm{Si}}$ denotes the energy of CD Si.
    Note that the curve obtained by the WGC KEDF overlaps with the curve of the LKT KEDF, while the curves obtained by KS-BLPS, HC, and \cpn{5} KEDFs nearly coincide.
    (b) Relative energy differences between HD Si and WZ semiconductors as obtained by OFDFT and KSDFT, where $E$ denotes the energy of specific structure, and $E_{\rm{Si}}$ denotes the energy of HD Si.
    Note that the curves obtained by KS-BLPS, HC, and \cpn{5} KEDFs are close, demonstrating that the latter two methods share similar accuracy with the KS-BLPS method.
    }
    \label{fig:E_diff}
\end{figure*}

\begin{table*}[b]
	\centering
	\caption{
MARE (Eq.~\ref{eq.mare}) of charge densities for the training set and the test set, as well as the average (Avg.) MARE among each set.
The training set consists of CD Si and nine ZB semiconductors, while the test set contains HD Si and nine WZ semiconductors.
MAREs are obtained by comparing various KEDFs (TF$\lambda$vW, LKT, WGC, HC, and CPN) in OFDFT to KS-BLPS results.
Note that nonconvergent results of the WGC KEDF in the test set have been excluded.
        }
	\begin{tabularx}{0.98\linewidth}{
			>{\raggedright\arraybackslash\hsize=1.33\hsize\linewidth=\hsize}X
			>{\centering\arraybackslash\hsize=0.97\hsize\linewidth=\hsize}X
			>{\centering\arraybackslash\hsize=0.97\hsize\linewidth=\hsize}X
			>{\centering\arraybackslash\hsize=0.97\hsize\linewidth=\hsize}X
			>{\centering\arraybackslash\hsize=0.97\hsize\linewidth=\hsize}X
			>{\centering\arraybackslash\hsize=0.97\hsize\linewidth=\hsize}X
			>{\centering\arraybackslash\hsize=0.97\hsize\linewidth=\hsize}X
			>{\centering\arraybackslash\hsize=0.97\hsize\linewidth=\hsize}X
			>{\centering\arraybackslash\hsize=0.97\hsize\linewidth=\hsize}X
			>{\centering\arraybackslash\hsize=0.97\hsize\linewidth=\hsize}X
			>{\centering\arraybackslash\hsize=0.97\hsize\linewidth=\hsize}X
			>{\centering\arraybackslash\hsize=0.97\hsize\linewidth=\hsize}X}
		\hline\hline
Training set		&CD Si 	&AlP 	&AlAs 	&AlSb 	&GaP 	&GaAs 	&GaSb 	&InP 	&InAs 	&InSb 	&Avg.\\
            \hline
TF$\rm{\lambda}$vW 	&14.22 	&18.90 	&18.76 	&17.62 	&18.40 	&18.53 	&18.09 	&21.65 	&21.57 	&20.45 	&18.82\\
LKT 				&28.29 	&33.04 	&32.81 	&32.91 	&33.44 	&33.49 	&33.20 	&36.76 	&36.52 	&36.78 	&33.72\\
WGC 				&4.66 	&14.81 	&12.78 	&8.06 	&13.05 	&11.39 	&6.93 	&19.92 	&17.48 	&10.85 	&11.99\\
HC 					&6.88 	&10.89 	&9.84 	&7.95 	&9.62 	&8.21 	&7.17 	&10.75 	&9.13 	&7.76 	&8.82\\
\cpn{1} 			&10.49 	&34.55 	&33.32 	&26.39 	&35.15 	&35.06 	&28.18 	&62.09 	&58.73 	&33.63 	&35.76\\
\cpn{3} 			&3.01 	&6.67 	&6.41 	&3.48 	&6.60 	&6.55 	&3.77 	&52.14 	&47.65 	&3.50 	&13.98\\
\cpn{5} 			&{0.84} 	&{1.38} 	&{1.08} 	&{2.88}	&{1.46} 	&{1.70} 	&{3.46} 	&{10.33} 	&{6.22} 	&{2.94} 	&{3.23}\\
            \hline
Test set &HD Si &AlP &AlAs &AlSb &GaP &GaAs &GaSb &InP &InAs &InSb &Avg.\\
            \hline
TF$\rm{\lambda}$vW &14.26 &19.22 &19.01 &17.83 &18.67 &18.77 &18.20 &21.82 &21.53 &20.01 &18.93\\
LKT &28.51 &33.55 &33.30 &33.38 &33.91 &33.96 &33.61 &37.40 &37.14 &37.53 &34.23\\
HC &6.87 &11.07 &10.03 &8.01 &9.68 &8.33 &7.31 &{11.09} &{9.41} &7.82 &8.96\\
\cpn{1} &26.97 &34.22 &33.27 &29.05 &34.73 &33.71 &29.89 &44.15 &42.54 &33.02 &34.15\\
\cpn{3} &17.18 &24.41 &23.54 &16.36 &24.55 &24.26 &17.44 &42.62 &38.13 &19.33 &24.78\\
\cpn{5} &{4.84} &{7.00} &{6.52} &{5.11} &{6.83} &{6.55} &{5.56} &15.24 &11.90 &{6.31} &{7.59}\\
		\hline\hline
	\end{tabularx}
	\label{tab:Density}
\end{table*}

In addition, the ground state energies of nine ZB semiconductors are calculated using various KEDFs, and the relative energy differences between CD Si and ZB semiconductors are displayed in figure~\ref{fig:E_diff}(a).
All tested KEDFs reproduced the energy ordering trends predicted by KSDFT among ten semiconductors.
However, the WGC and LKT KEDFs systematically overestimate the relative energies of ZB semiconductors, yielding MAREs of 9.7\% and 8.9\%, respectively.
The TF$\lambda$vW KEDF performs slightly better, exhibiting a MARE of 6.3\%, while the HC KEDF is the best-performing KEDF with a MARE of 1.4\%.
Additionally, the \cpn{1} and \cpn{3} KEDFs underestimate the relative energies, but the MAREs decrease significantly with an increasing number of channels (16.4\%, 5.9\%, and 2.6\% for $n = 1, 3, 5$ respectively).
The \cpn{5} KEDF outperforms the TF$\lambda$vW, LKT, and WGC KEDFs, approaching the accuracy of the HC KEDF.

Second, a similar test is performed on the test set, i.e., WZ semiconductors, to examine the transferability of the CPN KEDF.
Figure~\ref{fig:E_diff}(b) displays the relative energy differences between HD Si and WZ semiconductors.
We note that the WGC KEDF failed to achieve convergence in all ten structures of the test set, so we exclude the results of WGC KEDF below.
The LKT KEDF overestimates the relative energies again, with a MARE of 9.0\%, while the TF$\lambda$vW KEDF gives a slightly smaller MARE of 6.2\%.
As expected, the HC KEDF outperforms the two semilocal KEDFs and achieves a small MARE of 1.5\%.
Although HD Si and WZ semiconductors do not appear in the training set, all three CPN KEDFs can get convergence in these structures and reproduce the correct energy ordering trends predicted by KSDFT.
Once again, with more channels included, the accuracy of the CPN KEDFs increased rapidly, with MAREs of 12.6\%, 5.6\%, and 3.2\%, respectively.
As a result, we conclude that the \cpn{5} KEDF approaches the accuracy of the HC KEDF and exhibits satisfactory transferability by mixing the features gathered among different scales.

\SL{Third, to further assess the accuracy of the CPN KEDF, we calculated the ground state charge densities for 20 structures from both the training set and the test set}, and the MAREs are listed in table~\ref{tab:Density}.
In the training set, i.e., CD Si and ZB semiconductors, the semilocal TF$\lambda$vW and LKT KEDFs exhibit mean MARE of 18.82\% and 33.72\%, respectively, which are considerably larger than those obtained by the WGC and HC KEDFs (11.99\% and 8.82\%, respectively).
Impressively, the \cpn{5} KEDF yields a MARE of 3.23\%, which is much lower than all tested analytical KEDFs, including the HC KEDF.
As for the test set consisting of HD Si and WZ semiconductors, the MARE of \cpn{5} KEDF slightly increases to 7.59\% compared to the training set.
However, it is still lower than 8.96\% obtained by the HC KEDF, and much lower than 18.93\% and 34.23\% obtained by semilocal TF$\lambda$vW and LKT KEDFs, respectively.
Hence, the above results indicate both high accuracy and excellent transferability of the \cpn{5} KEDF.

\SL{In addition, we note that the unsatisfactory performance of the HC KEDF in predicting charge density may be attributed to the fact that its parameters, namely $\gamma$ and $\beta$ were optimized by fitting bulk properties obtained by KS-BLPS.
The bulk properties include the bulk modulus, equilibrium volume, and equilibrium energy, but not the charge density.
Furthermore, it is important to highlight that the optimal parameters for the HC KEDF are distinct for each specific structure tested in this study, which limits its practical value.~\cite{10B-Huang-nonlocal}
In contrast, the \cpn{5} KEDF describes these structures with the same set of parameters, thereby offering greater consistency and broader applicability.}

\begin{figure*}[t]
    \centering

    \includegraphics[width=0.95\linewidth]{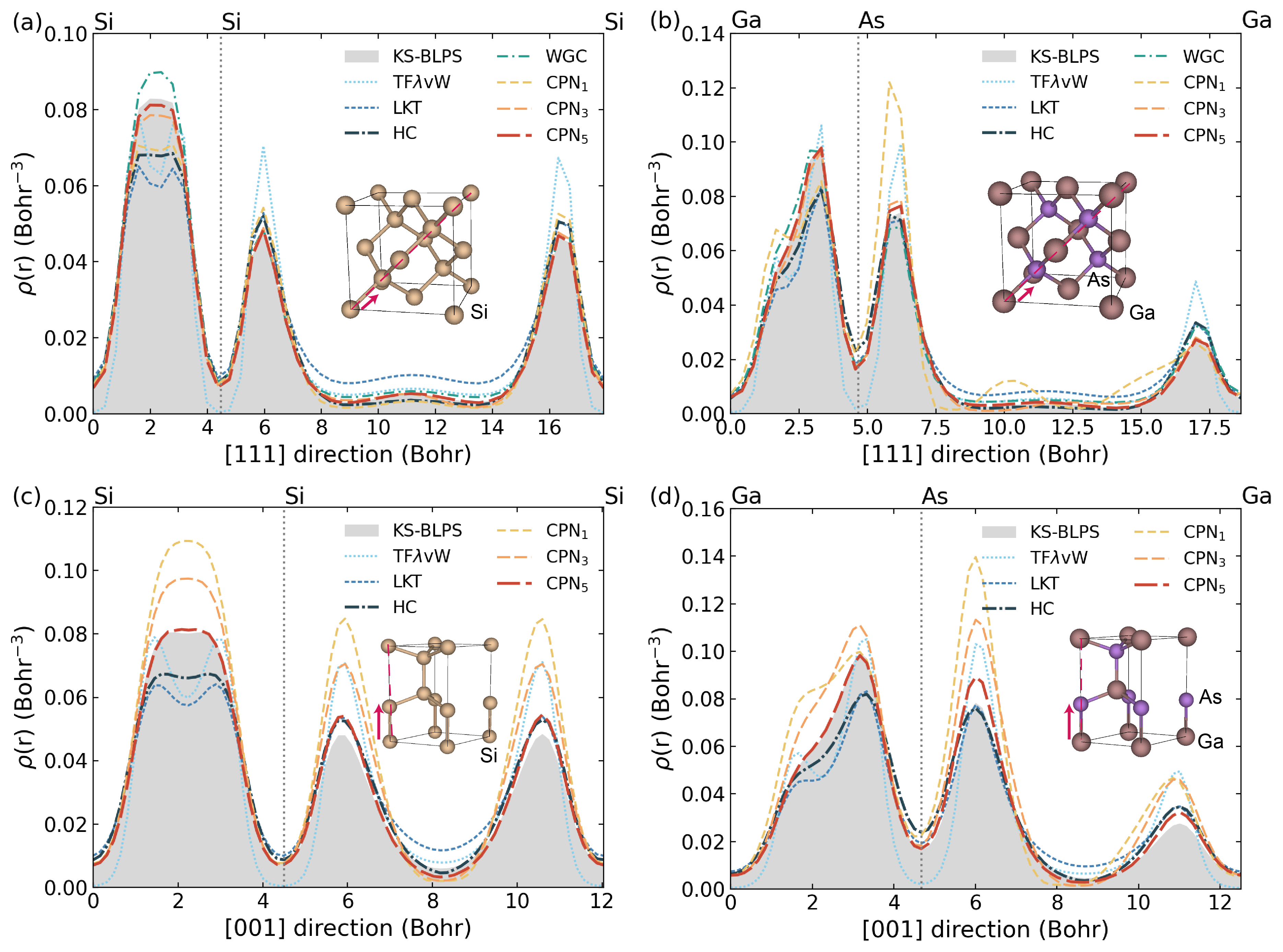}

    \caption{Charge densities $\rho(\mathbf{r})$ of four typical semiconductors, including
    (a) the CD structure of Si from the training set, (b) the ZB structure GaAs from the training set, (c) the HD structure Si from the test set, and (d) the WZ GaAs from the test set. The unit cells of CD Si and ZB GaAs, as well as the primitive cells of HD Si and WZ GaAs are illustrated in the insets. The [111] direction of CD Si and ZB GaAs, as well as the [100] direction of HD Si and WZ GaAs are illustrated by red arrows.
    }
    \label{fig:Density}
\end{figure*}

Figure~\ref{fig:Density} shows the charge densities of four typical semiconductors: CD Si and ZB GaAs from the training set, as well as HD Si and WZ GaAs from the test set.
First, all three CPN KEDFs can produce smooth charge densities.
This is a result of the smooth Pauli potential guaranteed by the Pauli potential term in the loss function and the nonlocal formula of Pauli potential, as derived in the SI of Ref.~\onlinecite{24B-Sun-mlof}.
Second, by comparing the charge densities obtained by \cpn{1}, \cpn{3}, \cpn{5} KEDFs, we can observe how the multi-channel architecture helps the ML model to capture the characteristics of semiconductors.
Moreover, the description of the covalent bond is a well-known hurdle for KEDFs.
Even the HC KEDF significantly underestimates the charge density in the covalent bond area.
However, the \cpn{5} KEDF can reproduce the covalent bond efficiently, both in the training set and the test set.
In conclusion, the above results emphasize the necessity of introducing the multi-channel architecture, which effectively helps the ML model handle the covalent bond.

\section{Conclusion}

In this work, we extended the MPN KEDF to semiconductor systems by integrating the features gathered among different scales using typical channels, and proposed the multi-channel MPN (CPN) KEDF.
The CPN KEDF inherits the advantages of the MPN KEDF, including the ‌incorporation of nonlocal information and adherence to three crucial physical constraints: the scaling law, the FEG limit, and the non-negativity of Pauli energy density.
%
The CPN KEDF was implemented in the ABACUS package.~\cite{16Li-CMS-ABACUS}

\SL{We conducted a systematic evaluation of various KEDFs on 20 semiconductors, comprising CD Si and nine ZB semiconductors (AlP, AlAs, AlSb, GaP, GaAs, GaSb, InP, InAs, InSb) in the training set, as well as HD Si and nine WZ semiconductors in the test set.
The ground state energy and charge density of these systems were calculated to assess the validity of the multi-channel architecture and to verify the accuracy, transferability, and stability of the CPN KEDFs.}
We have the following findings.
First, our results demonstrated that introducing multi-channel architecture is crucial for the CPN KEDF to capture the features of the electronic structure in semiconductors, for example, the localized electrons in the covalent bonds.
\SL{Second, the \cpn{5} KEDF, which incorporates five distinct channels, outperformed semilocal KEDFs and achieved an accuracy comparable to that of nonlocal KEDFs across all tested systems.}
Last, the \cpn{5} KEDF excellently reproduced the covalent bond structure, which is challenging for all tested analytical KEDFs.

Nevertheless, the CPN KEDFs are unable to reproduce the energy ordering of different phases of Si, such as CD and HD in the semiconductor phase, and $\beta$-tin, simple-cubic (sc), hexagonal-close-packed (hcp), body-centered-cubic (bcc), and face-centered-cubic (fcc) in the metallic phase.
Additionally, the CPN KEDFs are unable to reproduce the energy-volume curves of AlP, GaP, InP, AlAs, GaAs, and InAs in both ZB and WZ configurations.
These deficiencies can be attributed to two primary factors.
First, the absence of HD Si and ZB semiconductors at varying volumes from the training dataset hinders the accurate modeling of these systems.
Second, given that the kernel functions $\kernel{\lambda}$ utilized are spherically symmetric, the CPN KEDFs' descriptors are inherently incapable of capturing anisotropic information.
The incorporation of non-spherically symmetric kernel functions may address this limitation by providing a more nuanced description of the electronic structure.
Addressing these issues will be the major focus of future research endeavors.
\SL{
In summary, our proposed approach offers a new path to generate KEDFs that can handle the metal and semiconductor systems simultaneously.
Additionally, it provides a reference for the advancement of other ML-based functionals, such as exchange-correlation functionals.
}

\acknowledgements

The work of L.S. and M.C. was supported by the National Science Foundation of China under Grand No. 12074007 and No. 12122401. The numerical simulations were performed on the High-Performance Computing Platform of CAPT.

\section{APPENDIX}

\begin{figure*}[t]
    \centering

    \begin{subfigure}{0.47\textwidth}
    \centering
    \includegraphics[width=0.95\linewidth]{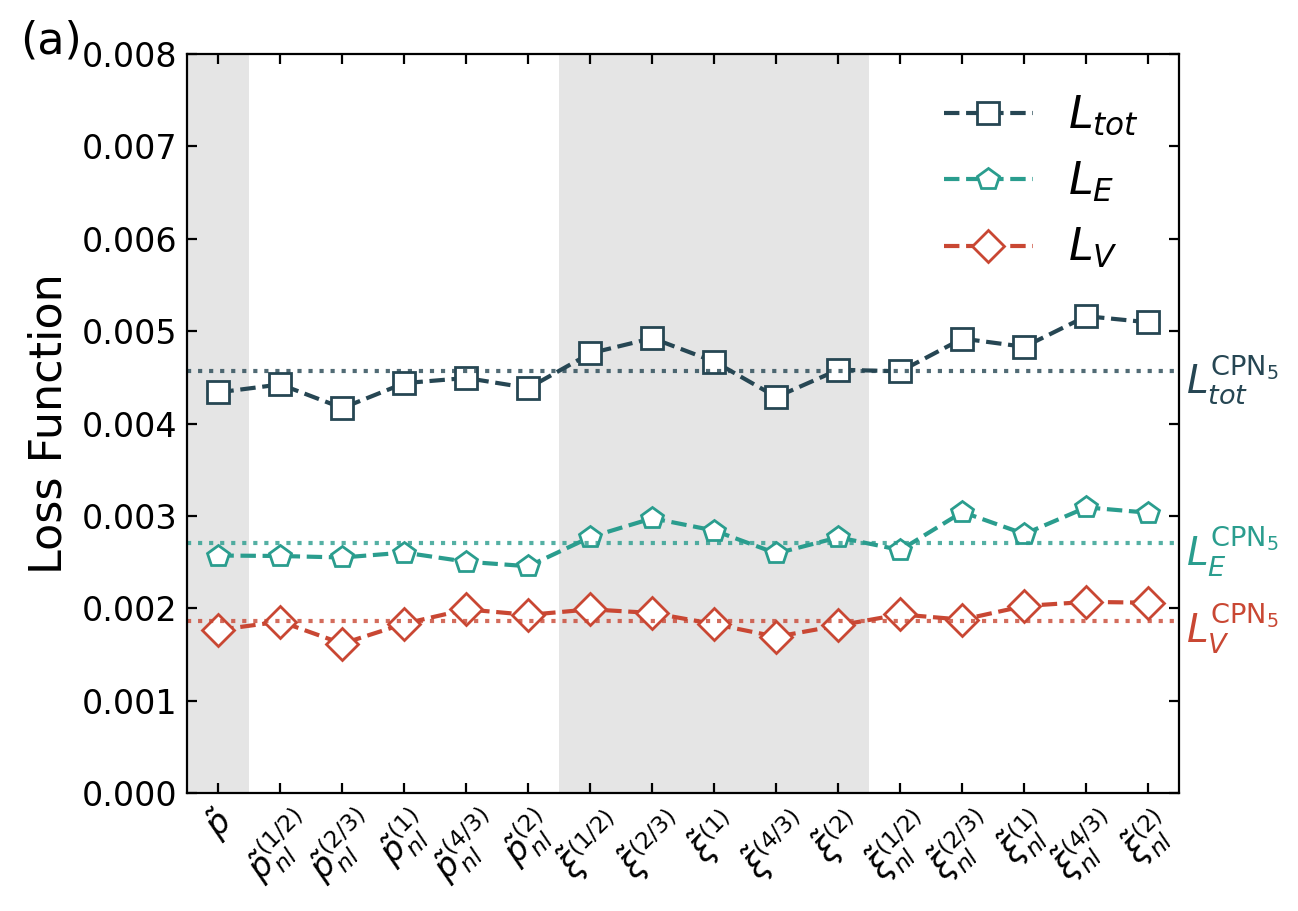}
    \label{fig:loss}
    \end{subfigure}
    \begin{subfigure}{0.52\textwidth}
    \centering
    \includegraphics[width=0.97\linewidth]{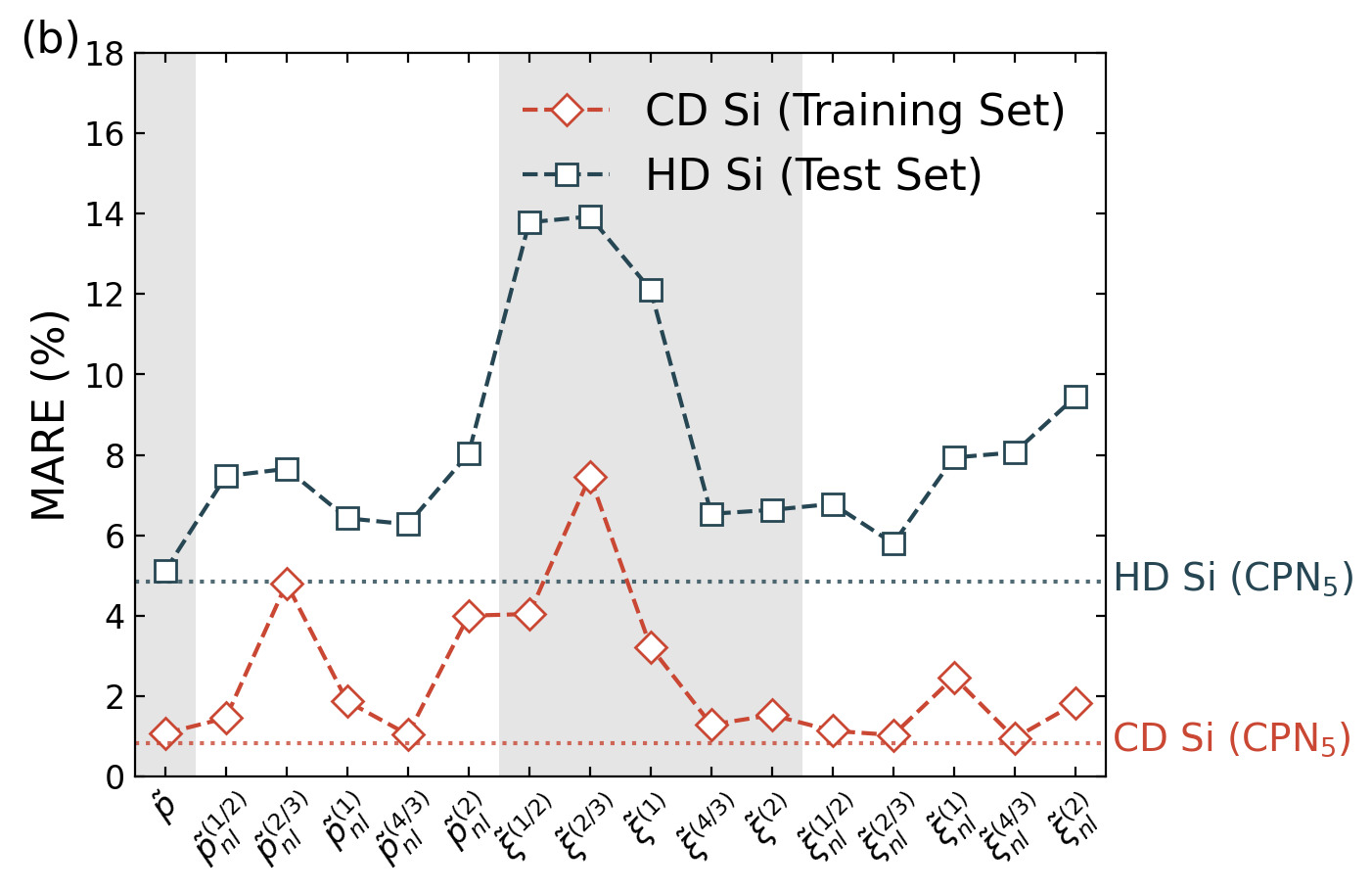}
    \label{fig:mare}
    \end{subfigure}

    \caption{\SL{
    (a) Total loss function $L_{\rm{tot}}$, alone with the energy term $L_{E}$ and the potential term $L_V$ for 16 distinct models obtained from the training set. Each model was developed by systematically omitting one specific descriptor from the set used in the \cpn{5} KEDF. The abscissa represents the omitted descriptor, and the loss functions of the \cpn{5} KEDF are illustrated by dashed lines.
    (b) MARE of the charge densities for CD Si from the training set and HD Si from the test set, as obtained by 16 distinct models. Each model was developed by systematically omitting one specific descriptor from the set used in the \cpn{5} KEDF. The abscissa represents the omitted descriptor, and the MARE values obtained by the \cpn{5} KEDF are illustrated by dashed lines.}
    }
    \label{fig:feature}
\end{figure*}

\SL{
To evaluate the importance of the selected descriptors, we have trained 16 distinct models,~\cite{ker2019image-fi, yang2022predicting-fi} each of which was developed by omitting one specific descriptor from the 16 descriptors used by the \cpn{5} KEDF. The total loss functions $L_{\rm{tot}}$ for these models are presented in Figure~\ref{fig:feature}(a).
As defined in Eq.~\ref{eq.loss}, the total loss function $L_{\rm{tot}}$ consists of three individual components, which include the energy term $L_E=\frac{1}{N}\sum_{\mathbf{r}}{ \left(\frac{F_\theta^{\rm{NN}}- F^{\rm{KS}}_{\theta}}{\bar{F}^{\rm{KS}}_{\theta}}\right)^2}$, the potential part $L_V=\frac{1}{N}\sum_{\mathbf{r}}{ \left(\frac{V_\theta^{\rm{CPN}} - V^{\rm{KS}}_{\theta}}{\bar{V}^{\rm{KS}}_{\theta}}\right)^2 }$, and the FEG part $L_{\rm{FEG}}=\left[F^{\rm{NN}}|_{\rm{FEG}}-\ln(e-1)\right]^2$.
Besides the total loss function, Figure~\ref{fig:feature}(a) also include the contributions from the energy term $L_E$ and the potential term $L_V$.
Note these values were averaged over the training set.
Additionally, the FEG component was not included in Figure~\ref{fig:feature}(a) due to its relatively minor contribution, being on the order of $10^{-9}$.
}
\SL{
Figure~\ref{fig:feature}(a) shows that the loss functions for all 16 models, along with their respective energy and potential terms, are close to those of the \cpn{5} KEDF. 
This similarity makes it challenging to directly assess the feature importance based solely on the loss function when only one descriptor is omitted.
}

\SL{
Furthermore, our experience indicates that minor variations in the loss function do not necessarily correlate with the accuracy and transferability of the model during self-consistent calculations. 
To address this, figure~\ref{fig:feature}(b) shows the MARE of the charge densities for CD Si from the training set and HD Si from the test set. 
We consider these MARE values as a more reliable criterion for evaluating feature importance.}
\SL{
By analyzing the MARE of CD Si, we have determined that $\Tilde{\xi}^{(2/3)}, \Tilde{p}^{(2/3)}_{\rm{nl}}, \Tilde{\xi}^{(1/2)}, \Tilde{p}^{(2)}_{\rm{nl}}, \Tilde{\xi}^{(1)}$ are the five most critical features to determine the accuracy of the model. Additionally, by comparing the MARE of HD Si, we conclude that $\Tilde{\xi}^{(2/3)}, \Tilde{\xi}^{(1/2)}, \Tilde{\xi}^{(1)}, \Tilde{\xi}^{(2)}_{\rm{nl}}, \Tilde{\xi}^{(4/3)}_{\rm{nl}}$ are the five most important features for the model's transferability. 
We note that these descriptors encompass all five channels, underscoring the crucial role of the multi-channel architecture in both the accuracy and transferability of the model.
}

\SL{
Moreover, if additional descriptors are removed, the loss function increases significantly.
For example, deducting $\{\Tilde{p}^{(4/3)}_{\rm{nl}}, \Tilde{\xi}^{(4/3)}, \Tilde{\xi}^{(4/3)}_{\rm{nl}}, \Tilde{p}^{(2/3)}_{\rm{nl}}, \Tilde{\xi}^{(2/3)}, \Tilde{\xi}^{(2/3)}_{\rm{nl}}\}$ transforms the \cpn{5} KEDF into the \cpn{3} KEDF, resulting in an increase in the loss function from 0.0046 to 0.0073.
A further deduction of the set $\{\Tilde{p}^{(2)}_{\rm{nl}}, \Tilde{\xi}^{(2)}, \Tilde{\xi}^{(2)}_{\rm{nl}}, \Tilde{p}^{(1/2)}_{\rm{nl}}, \Tilde{\xi}^{(1/2)}, \Tilde{\xi}^{(1/2)}_{\rm{nl}}\}$ to obtain the \cpn{1} KEDF leads to an even greater increase in the loss function to 0.0217.
These observations highlight the significant importance of the descriptors from these channels in maintaining the model's performance.
}

\bibliography{ML-KEDF}
\end{document}